\documentstyle[twoside,fleqn,espcrc2]{article}

\newcommand{\AmS}{{\protect\the\textfont2
  A\kern-.1667em\lower.5ex\hbox{M}\kern-.125emS}}

\title{Two-dimensional Yang-Mills theory: perturbative and
instanton contributions, and its relation to QCD in higher dimensions}

\author{A. Bassetto\address{Dipartimento di Fisica ``G. Galilei'' 
and INFN, Sezione di Padova, \\ 
Via Marzolo 8, 35131 Padua, Italy}%
\thanks{Plenary session talk at the Meeting ``Constrained
dynamics and quantum gravity 99'', Villasimius (Sardinia-Italy)
September 13-17, 1999.}}

\begin{document}

\begin{abstract}
Two different scenarios (light-front and equal-time) are possible
for Yang-Mills theories in two dimensions. The exact $\bar q q$-potential
can be derived in perturbation theory starting from the light-front vacuum,
but requires essential instanton contributions in the equal-time
formulation. In higher dimensions no exact result is available and,
paradoxically, only the latter formulation (equal-time) is
acceptable, at least in a perturbative context.
\end{abstract}

\maketitle

\section{INTRODUCTION}

The non-perturbative structure of non-abelian quantum gauge theories
is still a challenging topic in spite of a large amount of efforts in
the recent literature. Whereas perturbation theory provides a
well-established frame to describe the weak-coupling
regime, quantitative predictions for the strong-coupling behaviour 
are extremely hard to be obtained.
Though some non-perturbative
features are thought to be transparent, a consistent framework in the
continuum is still lacking.

Such problems
have often been tackled in the simplified context of
two-dimensional gauge theories ($YMT_2$), taking advantage of the lattice
solutions \cite{migdal}. As far as the continuum is concerned,
in two dimensions the theory looks seemingly trivial when quantized
in the light-cone gauge (LCG) $A_-\equiv \frac{A_0-A_1}{\sqrt 2}=0$ . 
As a matter of fact, in the absence of dynamical
fermions, it looks indeed free, being described by a Lagrangian quadratic
in the fields.

Still topological degrees of freedom occur if the theory is put
on a (partially or totally) compact manifold, whereas the simpler 
behavior on the plane enforced by the LCG condition 
entails a severe worsening in its infrared structure.
These features are related aspects of the same basic issue: even in two
dimensions ($D=2$) the theory contains non trivial dynamics, 
as immediately suggested
by other gauge choices as well as by perturbative calculations of
gauge invariant quantities, typically of Wilson loops\cite{noi1}.
We can say that, in LCG, dynamics gets hidden
in the very singular nature of correlators at large distances
(IR singularities). 

In order to fully appreciate this point and the controversial aspects
related to it, it is useful to review briefly the 't Hooft's
model for $QCD_2$ at large $N$, $N$ being the number of colours
\cite{thooft}.
In LCG no self-interaction occurs for the gauge fields; in the large-$N$
limit planar diagrams dominate and the $q\bar q$ interaction is
mediated by the exchange
\begin{equation}
\label{exch}
{\cal D}(x)=-\frac{i}{2}|x^-|\,\delta(x^+),
\end{equation}
which looks instantaneous if $x^+$ is considered a {\it time} variable.
Eq.(\ref{exch}) is the Fourier transform of the
quantity
\begin{equation}
\label{four}
\tilde{\cal D}(k)=\frac{1}{k_-^2},
\end{equation}
the singularity at $k_-=0$ being interpreted as a Cauchy principal value.
Such an expression in turn can be derived by quantizing the theory on
the light-front (at equal $x^+$), $A_+$ behaving like a constraint
\cite{libro,noi1}.

The full set of planar diagrams can easily be summed, leading to a 
beautiful pattern of $q\bar q$-bound states with squared masses
lying on rising Regge trajectories. This was the first evidence, 
to our knowledge, of a {\it stringy} nature of $QCD$ in its confining 
regime, reconciling dual models with a partonic field theory.

After this pioneering investigation, many interesting papers followed
't Hooft's approach, blooming into the more recent achievements of
$QCD_2$.

\smallskip

Still, if the theory within the same gauge choice is canonically
quantized {\it at equal times}, a different expression is obtained for
the exchange in eq.(\ref{exch})
\begin{equation}
\label{caus}
{\cal D}_{c}(x)=\frac{1}{2\pi}\,\frac{x^-}{-x^++i\epsilon x^-},
\end{equation}
and its Fourier transform
\begin{equation}
\label{fcaus}
\tilde {\cal D}_{c}(k)=\frac{1}{(k_-+i\epsilon k_+)^2},
\end{equation}
can now be interpreted as a {\it causal} Feynman propagator \cite{noi1}.

This expression, first proposed by Wu \cite{wu}, is nothing but
the restriction at $D=2$ of the prescription for the LCG vector
propagator in four dimensions suggested by Mandelstam \cite{mandel}
and Leibbrandt \cite{leibb} (ML), and derived in ref.\cite{noi}
by equal-time canonical quantization of the theory.

In dimensions higher than two, where ``physical'' degrees of freedom
are switched on (transverse ``gluons''), this causal prescription is
the only acceptable one; indeed causality is mandatory in order to
get correct analyticity properties, which in turn are the basis of
any consistent renormalization program\cite{noi2}.
It has been shown in perturbative calculations \cite{bass} that agreement 
with Feynman gauge results can only be obtained if a causal propagator is used 
in LCG.

This causal behaviour is induced by the 
propagation of unphysical degrees of freedom (probability ghosts), 
which can be expunged from the ``physical'' Hilbert space,
but still contribute in intermediate lines 
as timelike
photons do in the QED Gupta-Bleuler quantization.
The presence of those ghosts will have far-reaching consequences in our 
subsequent considerations.

\smallskip

When eq.(\ref{fcaus}) is used in summing the very same set of planar 
diagrams considered by 't Hooft, no rising Regge trajectories
are found in the spectrum of the $q\bar q$-system. The bound-state
integral equation looks difficult to be solved; early approximate
treatments \cite{webb,mitra} as well as a more detailed recent study 
\cite{shuva}
indicate the presence of a massless solution, with a fairly obscure 
interpretation, at least in this context. Confinement seems lost.

Then, how can it be that the causal way to treat the infrared (IR) 
singularities,
which is mandatory in higher dimensions, leads to a disastrous result
when adopted at $D=2$ ? In order to get an answer we turn to an 
investigation concerning the $q\bar q$-potential.

\section{THE WILSON LOOP}

A very convenient gauge invariant way of looking at the $q\bar q$-potential 
is by first considering a rectangular Wilson loop $\gamma$ with
sides parallel to a spatial direction and to the time direction
\begin{equation}
\label{wilson}
{\cal W}_{\gamma} = {1\over N} \langle 0| {\rm Tr}\left[ {\cal T}{\cal P} 
{\rm exp} \left( ig \oint_\gamma dx^\mu A_\mu (x) \right)\right]
|0\rangle, 
\end{equation}
the symbols ${\cal T}$ and ${\cal P}$ denoting temporal ordering of
operators and colour ordering, respectively.

The contour $\gamma$ can be parametrized as
\begin{eqnarray}
\gamma_1 &:& \gamma_1^\mu (s) = (sT, L)\ ,\nonumber\\
\gamma_2 &:& \gamma_2^\mu (s) = (T,-sL)\ ,\nonumber\\
\gamma_3 &:& \gamma_3^\mu (s) = (-sT, -L)\ , \nonumber\\
\gamma_4 &:& \gamma_4^\mu (s) = (-T, sL)\ , \ \ \qquad -1 \leq s \leq 1, 
\label{path}
\end{eqnarray}
describing a  (counterclockwise-oriented) rectangle
centered at the origin of the plane ($x^1,x^0$),
with sides of length $(2L,2T)$, respectively.

It is well known that the Wilson loop we have hitherto introduced
can be thought to describe the interaction of a couple of static 
$q\bar q$ at the distance $2L$ from each other. If we denote by
\begin{equation}
\label{stringa}
{\cal M}(y,x;\Gamma)=\bar q(y)E[\Gamma]q(x)
\end{equation}
the mesonic string operator, $E[\Gamma]$ representing the gluon phase
factor along the path $\Gamma$ connecting the couple, we can consider
the overlap between the states $\bar q q$ at the times $t=-T$ and
$t=T$, that is the amplitude
\begin{equation}
\label{overlap}
M(L,T)=\langle 0|{\cal M}^{\dagger}(2L,T){\cal M}(2L,-T)|0\rangle.
\end{equation} 
If we insert a complete set of eigenstates $|\Phi_{n}\rangle$ which
diagonalizes the Hamiltonian of the system $\bar q q$ at the distance 
$2L$ with eigenvalues ${\cal E}_{n}(L)$,
we easily obtain
\begin{equation}
\label{spettro}
M(L,T)=\sum_{n}|\langle 0|{\cal M}(2L,0)|\Phi_{n}\rangle|^2 
\exp(2i{\cal E}_{n}T).
\end{equation} 
We can turn to the Euclidean formulation replacing $T$ with $iT$. If we 
denote by ${\cal E}_0(L)$ the ground state energy of the system, we get
\begin{eqnarray}
\label{ground}
&&M(L,T)=\exp(-2{\cal E}_{0}T)\nonumber \\
&&\int_{{\cal E}_{0}}^{\infty}
d{\cal E}\,\rho(L,{\cal E})\,\exp[-2T({\cal E}-{\cal E}_0)].
\end{eqnarray} 
Unitarity requires the spectral density
$\rho(L,{\cal E})$ to be a non-negative measure (see 
eq.(\ref{spettro})). Then $M(L,T)$ is positive and the coefficient
of the exponential factor $\exp(-2{\cal E}_{0}T)$ is a non-increasing
function of $T$.

\smallskip

In the limit $T\to \infty$ one can show \cite{band} that the
Wilson loop ${\cal W}_{\gamma}$ is related to $M(L,T)$ by a
threshold factor 
\begin{equation}
\label{thresh}
{\cal W}_{\gamma}\simeq \exp(4mT)\,M(L,T),
\end{equation}
m being the static quark mass. In this way we can define the 
$q\bar q$-potential $${\cal V}(L)={\cal E}_0(L)-2m.$$
If the theory confines, ${\cal V}(L)$ is an increasing function of
the distance $L$; if at large distances the increase is linear in $L$,
namely $${\cal V}(L)\simeq 2 \sigma L,$$ we obtain an area-law behaviour 
for the
leading exponent, characterized by a string tension $\sigma$. 

For $D>2$ perturbation theory is unreliable in computing
the true spectrum of the $q\bar q$-system. However, when combined
with unitarity, it puts an
intriguing constraint on the $q\bar q$-potential. To realize this point,
let us consider the formal expansion
\begin{equation}
\label{expand}
{\cal V}(L)=g^2\, {\cal V}_1(L)+g^4\, {\cal V}_2(L)+\cdots,
\end{equation}
$g$ being the QCD coupling constant.
When inserted in the expression $\exp[-2{\cal V}(L)T]$, it gives 
\begin{eqnarray}
\label{loopex}
&&\exp[-2T{\cal V}]= 1-2T\bigl[g^2\, {\cal V}_1+g^4\, 
{\cal V}_2+\cdots\bigr]\nonumber \\
&&+2T^2\bigl[g^4\,{\cal V}_1^2+\cdots\bigr]+\cdots.
\end{eqnarray}
At ${\cal O}(g^4)$, the coefficient
of the leading term at large $T$ should be half the square of the term
at ${\cal O}(g^2)$. This constraint has often been used as a check
of (perturbative) gauge invariance \cite{libro}.

Therefore, if we denote by $C_{F(A)}$ the quadratic Casimir expression for the
fundamental (adjoint) representation
of $SU(N)$ and remember that ${\cal V}_1$ is proportional to $C_{F}$,
at ${\cal O}(g^4)$ the term with the coefficient $C_{F}C_{A}$
should be subleading in the large-$T$ limit with respect to the
Abelian-like term, which is proportional to $C_{F}^2$.

Such a calculation at ${\cal O}(g^4)$ for the loop ${\cal W}_{\gamma}$
has been performed using Feynman gauge in ref.\cite{belluo}, with  
the number of space-time dimensions
larger than two ($D>2$). The parameter D acts also as a (gauge invariant)
dimensional regulator.

The result depends on the area ${\cal A}=L\,T$ and on the dimensionless
ratio $\beta=\frac{L}{T}$. The ${\cal O}(g^2)$-term is obviously
proportional to $C_{F}$; at ${\cal O}(g^4)$ we find 
that the non-Abelian term is indeed
subleading
\begin{equation}
\label{sublead}
T^2\,{\cal V}^{na}\propto C_{F}C_{A}{\cal A}^2\, T^{4-2D},
\end{equation}
as expected.

Therefore agreement with exponentiation holds and the validity
of previous perturbative tests of gauge invariance in higher
dimensions (see ref.\cite{libro}) 
is vindicated. This rather simple way of
realizing the exponentiation at $D>2$ might have a deeper justification
as well as far-reaching consequences.

The limit of our result when $D\to 2$ is {\it finite} and depends only
on ${\cal A}$, as expected on the basis of the invariance of the theory
in two dimensions under area-preserving diffeomorphisms. However the
non-Abelian term is no longer subleading in the limit $T\to \infty$,
as it is clear from eq.(\ref{sublead}); we get instead \cite{belluo}
\begin{equation}
\label{lead}
2T^2\,{\cal V}^{na}= C_{F}C_{A}\frac{{\cal A}^2}{16\pi^2}(1+\frac{\pi^2}{3}).
\end{equation}

We conclude that the limits $T\to \infty$ and $D\to 2$ {\it do not commute}.

This result is confirmed by a calculation of ${\cal W}_{\gamma}$ performed
in LCG with the ML prescription for the vector propagator \cite{colf,bell}. 
At odds with Feynman gauge where the vector
propagator is not a tempered distribution at $D=2$, in LCG the calculation
can also be performed directly in two space-time dimensions. The result
one obtains does {\it not} coincide with eq.(\ref{lead}). One gets instead
\begin{equation}
\label{lead2}
2T^2\,{\cal V}^{na}= C_{F}C_{A}\frac{{\cal A}^2}{48}.
\end{equation}

The extra term in eq.(\ref{lead}) originates from the self-energy correction
to the vector propagator. In spite of the fact that the triple vector
vertex vanishes in two dimensions in LCG, the self-energy correction 
does not.
What happens is that the vanishing of triple vertices when $D\to 2$
is exactly compensated by the loop integration singularity
at $D=2$ leading, eventually, to a finite result. We would like to
stress that this ``anomaly-like'' contribution is not a pathology 
of LCG; precisely it is needed
to get agreement with the Feynman 
gauge result.

Perturbation theory is {\it discontinuous} at $D=2$.
 
We conclude that the perturbative result, no matter
the gauge one adopts, conflicts with unitarity in two dimensions.
We recall that any acceptable gauge (including the causal formulation of
LCG!) entails the
presence of (probability) ghosts at $D>2$ (see ref.\cite{libro}). At $D=2$
in LCG these ghosts are the only surviving degrees of freedom in 
perturbative YMT
\cite{noi1};
therefore a possible violation of unitary is hardly surprising, although 
the reason why ghosts are so dangerous precisely in two dimensions
is not yet fully clear.

Taking advantage of the invariance under area-preserving diffeomorphisms
in dimensions $D=2$, Staudacher and Krauth \cite{sk} were
able to resum 
the perturbative series at all orders in the coupling constant $g$ 
in LCG within the causal formulation,
thereby generalizing our ${\cal O}(g^4)$ result (eq.(\ref{lead2})). 
In the Euclidean formulation, which is possible as the causal
propagator can be Wick-rotated, and with a particular choice of the
contour (a circumference), 
they show that the colour factors decouple
from geometry and can be summed by the simple matrix integral
\begin{eqnarray}
\label{crauti}
&&{\cal W}_{\gamma}({\cal A})=\frac{1}{\cal Z}\int {\cal D}F\,
\exp(-\frac{1}{2}{\rm Tr}F^2)\nonumber \\
&&\frac{1}{N}{\rm Tr}\exp(igF\sqrt{\frac{\cal A}{2}}),
\end{eqnarray}
where, for $U(N)$, ${\cal D}F$ denotes the flat integration measure 
on the space of constant Hermitian $N\times N$ matrices and
${\cal W}_{\gamma}(0)=1 $.
They get
\begin{equation}
\label{krauth}
{\cal W}_{\gamma}({\cal A})
=\frac{1}{N}\exp\Big[- \frac { g^2\cal A}{4}\Big]L^{(1)}_{N-1}
\Bigl(\frac{g^2 {\cal A}}{2}\Bigr),
\end{equation}
the function $L^{(1)}_{N-1}$ being a  Laguerre polynomial.

This result can be easily generalized to a loop winding $n$-times
around the countour
\begin{equation}
\label{risultatino}
{\cal W}_{\gamma}=\frac{1}{N}\exp\left[-\frac{g^2 {\cal A} \,n^2}4\right]\,
L_{N-1}^{(1)}\left(\frac{g^2 {\cal A} \,n^2}{2} \right).
\end{equation}
 
From eq.(\ref{krauth}) one immediately realizes that, 
for even values of $N$, the result is no
longer positive in the large-$T$ limit. Moreover in the 't Hooft's
limit $N \to \infty$ with $g^2 N=2\hat g^2$ fixed, the string
tension vanishes and eq.(\ref{krauth}) becomes
\begin{equation}
\label{bessel}
{\cal W}_{\gamma} \to \frac{1}{\sqrt{\hat g^2 {\cal A}}}
J_1(2\sqrt{\hat g^2 {\cal A}}),
\end{equation}
$J_1$ being the usual Bessel function.

Confinement is lost.

This explains the failure of the Wu's approach in getting a bound state
spectrum lying on rising Regge trajectories in the large-$N$ limit.
The planar approximation is inadequate when the $q\bar q$-interaction
is described in a causal way.

\smallskip

However in LCG 
the theory can also be quantized on the {\em light-front} (at equal $x^{+}$);
with such a choice, in pure YMT and just in two dimensions,
no dynamical degrees of freedom occur as the non
vanishing component of the vector field does not propagate,
but rather gives rise to an instantaneous (in $x^{+}$) Coulomb-like
interaction (see eq.(\ref{exch})). There are no problems with 
causality and renormalization is no longer a concern.

If the Wilson loop ${\cal W}_{\gamma}$ is perturbatively computed
using expression (\ref{exch}), only planar diagrams contribute for
any value of $N$, thanks to the ``instantaneous'' nature of such
an exchange; the perturbative series can be easily resummed,
leading to the result (for imaginary time)
\begin{equation}
\label{area}
{\cal W}_{\gamma}({\cal A})
=\exp\Big[- \frac { g^2N\cal A}{4}\Big],
\end{equation}
to be compared with eq.(\ref{krauth}).

Not only is this result in complete agreement with the exponentiation 
required by unitarity; it also exhibits, in the 't Hooft's limit
$N \to \infty$ with $g^2 N=2\hat g^2$ fixed, confinement with a finite string
tension $\sigma=\frac{\hat g^2}{2}$.

This explains the success of 't Hooft's approach in computing the spectrum
of the $q\bar q$ bound states.

The deep reason of this good behaviour lies in the absence of ghosts
in this formulation; however we should stress again that it cannot
be derived in a smooth way from any acceptable gauge choice in
higher dimensions $(D>2)$. Moreover the confinement exhibited at this stage
is, in a sense, trivial, since it occurs also in
the Abelian case $U(1)$, namely $QED_2$.

We end up with two basically different results for the {\it same} model
and with the {\it same} gauge choice (LCG), according to the different ways
in which IR singularities are regularized.
Moreover we are confronted with the following paradox:
the prescription which is mandatory in dimensions $D>2$ is the
one which fails at $D=2$. What is the meaning (if any) of eq.(\ref{krauth})?

\section{THE GEOMETRICAL APPROACH}

In order to understand this point, it is worthwhile to study the problem
on a compact two-dimensional manifold; possible IR
singularities will be automatically regularized in a gauge invariant way.
For simplicity, we choose
the sphere $S^2$. We also consider the slightly simpler
case of the group $U(N)$ (the generalization to $SU(N)$ is straightforward)
\cite{gross}.  
On $S^2$ we consider a smooth non self-intersecting closed contour 
and a loop winding around it a number $n$ of times.
We call $A$ the total area of the sphere,
which eventually will be sent to $\infty$, whereas ${\cal A}$ will be
the area ``inside'' the loop we keep finite in this limit.

Our starting point is the well-known heat-kernel 
expressions \cite{migdal} of a non self-intersecting Wilson loop
for a pure $U(N)$ YMT on a sphere $S^2$ with area $A$
\begin{eqnarray}
\label{wilsonz}
&&{\cal W}_{n}(A,{\cal A})={1\over {\cal Z}(A)N}
\sum_{R,S} d_{R}d_{S}\nonumber \\
&&\exp\left[-{{g^2 {\cal A}}\over 4}C_2(R)-{{g^2 (A-{\cal A})}\over 4}
C_2(S)\right]\nonumber \\
&&\times \int dU {\rm Tr}[U^{n}]\chi_{R}(U) \chi_{S}^{\dagger}(U),
\end{eqnarray}
$d_{R\,(S)}$ being the dimension of the irreducible representation $R(S)$ of
$U(N)$; $C_2(R)$ ($C_2(S)$) is the quadratic Casimir expression, 
the integral in (\ref{wilsonz}) is over the
$U(N)$ group manifold while $\chi_{R(S)}$ is the character of the group
element $U$ in the $R\,(S)$ representation.
In the sequel the partition function ${\cal Z}(A)$
will be rescaled by absorbing suitable factors 
without changing notation. Normalization
will always be recovered by setting ${\cal W}_{n}(A,0)=1.$

We write eq.(\ref{wilsonz}) explicitly for $N>1$ and $n>0$ in the form
\begin{eqnarray}
\label{wilsonp}
&&{\cal W}_{n}(A,{\cal A})=\frac{1}{{\cal Z}(A)}
\sum_{m_i=-\infty}^{+\infty}\Delta(m_1,...,m_N)\nonumber\\
&&\times\Delta(m_1+n,m_2,...,m_N)\nonumber \\
&&\times \exp\left[-\frac{g^2A}{4}\sum_{i=1}^N (m_i)^2
 \right]\nonumber \\
&&\times \exp\left [-\frac{g^2 n}{4}(A-{\cal A})(n+2m_1)\right].
\end{eqnarray}
We have described the generic irreducible representation by means
of the set of integers $m_{i}=(m_1,...,m_{N})$, related to the
Young tableaux, in terms of which
we get
\begin{eqnarray}
\label{casimiri}
C_2(R)&=&\frac{N}{12}(N^2-1)+\sum_{i=1}^{N}(m_{i}-\frac{N-1}{2})^2,\nonumber
\\
d_{R}&=&\Delta(m_1,...,m_{N}).
\end{eqnarray}
$\Delta$ is the Vandermonde determinant and
the integration in eq.(\ref{wilsonz})
has been performed explicitly, using the well-known formula for the 
characters in terms of the set $m_{i}$ and taking symmetry into account.

From eq.(\ref{wilsonp}) it is possible to derive, for $n=1$ and 
in the large-$A$ 
decompactification limit, precisely the expression 
in eq.(\ref{area}) we obtained by resumming the perturbative series
in the 't Hooft's approach \cite{boul}. 
This is a remarkable result as it has now been
derived in a purely geometrical way without even fixing a gauge.

Actually, in the decompactification limit $A\to \infty$ at fixed
${\cal A}$, from eq.(\ref{wilsonp}) one can derive the following expression
for any value of $n$ and $N$ \cite{bgv}
\begin{eqnarray}
\label{wilmorenice}  
&&{\cal W}_{n}({\cal A};N)=\frac{1}{n N} \, \exp\Bigl[-\frac{g^2 {\cal A}}{4}
\, n(N+n-1)\Bigr]\nonumber \\
&&\times\!\!\sum_{k=0}\!\frac{(-1)^k\,\Gamma(N+n-k)}{k!\,\Gamma(N-k)\Gamma(n-k)}
\exp\Bigl[\frac{g^2 {\cal A}}{2}\,n\,k \Bigr].
\end{eqnarray}
 
We notice 
that when $n>1$ the simple abelian-like exponentiation is lost. In other 
words the theory starts feeling its non-abelian nature as the appearance 
of different ``string tensions'' makes clear.
The winding number $n$ probes its colour content.
The related light-front vacuum, although simpler than the one in the
equal-time quantization, cannot be considered trivial any longer.
The above result
does not seem related to any simple-minded reduction $U(N) \sim U(1)^N$,
as suggested by the abelianization of the theory in axial gauges.

Eq.(\ref{wilmorenice}) exhibits an interesting symmetry between $N$ and $n$. 
More precisely, we have that
\begin{eqnarray}
\label{dual}
{\cal W}_{n}({\cal A};N)&=&{\cal W}_{N}(\tilde{\cal A};n),\\
\tilde{\cal A} &=& \frac {n}{N} \,{\cal A} \,,\nonumber
\end{eqnarray}
a relation that is far from being trivial, involving an unexpected
interplay between the geometrical and the algebraic structure of the
theory \cite{bgv}.

Looking at eq.(\ref{dual}), the abelian-like exponentiation for $U(N)$
when $n=1$ appears to be related to the $U(1)$ loop with $N$
windings, the ``genuine'' triviality of Maxwell theory providing the
expected behaviour for the string tension. Moreover we notice the
intriguing feature that the large-$N$ limit (with $n$ fixed) is
equivalent to the limit in which an infinite number of windings is
considered with vanishing rescaled loop area. Alternatively, this 
rescaling could be thought to affect the
coupling constant $g^2 \to \frac{n}{N} g^2$.

From eq.(\ref{wilmorenice}), in the limit $N\to \infty$, one can recover
the Kazakov-Kostov result \cite{kaza}
\begin{equation}
\label{vecchia}
{\cal W}_{n}({\cal A};\infty) = \frac1n 
L^{(1)}_{n-1}(\frac{\hat{g}^2 {\cal A}n}{2}) \,
\exp \Bigl[ -\frac{\hat{g}^2 {\cal A} n}4\Bigr].
\end{equation}

Now, using eq.(\ref{dual}) we are able to perfom another limit, namely $n\to
\infty$ with fixed $n^2\,{\cal A}$  
\begin{eqnarray}
\label{granden}
&& \lim_{n\to\infty} {\cal W}_{n}({\cal A};N)  =
\frac 1N \,
L^{(1)}_{N-1}\Bigl( g^2 {\cal A}\,n^2 /2 \Bigr)\nonumber \\
&& \exp \Bigl[ -\frac{g^2 {\cal A}\, n^2}4\Bigr]  \,.
\end{eqnarray}

We remark that this large-$n$ result reproduces the resummation of the
perturbative series (for any $n$) (eq.(\ref{risultatino})) in the {\it causal}
formulation of the theory. This is not a coincidence, rather it provides
a clue to understand its deep meaning.

\smallskip

We go back to the exact expressions we have found on the sphere 
for the Wilson loop (eq.\ref{wilsonp}).
As first noted by Witten \cite{Witte}, it is possible to
represent ${\cal W}_{n}(A,{\cal A})$ (and consequently ${\cal Z}(A)$) 
as a sum over
instable instantons, where each instanton contribution is 
associated to a finite,
but not trivial, perturbative expansion. The easiest way to see it, is 
to perform a Poisson resummation \cite{case,gross}
\begin{eqnarray}
\label{poisson}
&&\sum_{m_{i}=-\infty}^{+\infty}\!F(m_1,...,m_{N})=
\sum_{f_{i}=-\infty}^{+\infty}\!\tilde{F}(f_1,...,f_{N}),\nonumber\\
&&\tilde{F}(f_1,...,f_{N})=\int_{-\infty}^{+\infty}dz_1...dz_{N}
F(z_1,...,z_{N})\nonumber \\
&&\times \exp \left[2\pi i(z_1 f_1+...+z_{N}f_{N})\right]
\end{eqnarray}
in eq.(\ref{wilsonp}).
One gets
\begin{eqnarray}
\label{istanti}
&&{\cal W}_{n}(A,{\cal A})=\frac{1}{{\cal Z}(A)}\exp\left[\frac{g^2n^2
(A-2{\cal A})^2}{16 A}\right]\nonumber \\
&&\sum_{f_{i}=-\infty}^{+\infty}
\exp\left[-S_{inst}(f_{i})\right]W(f_1,...,f_{N})\nonumber\\
&&\times\exp\left[-2 \pi in f_{1}\frac{A-{\cal A}}{A}\right],
\end{eqnarray}
where
\begin{equation}
\label{quantita`}
S_{inst}(f_{i})=\frac{4\pi^2}{g^2 A}\sum_{i=1}^{N}f_{i}^2,
\end{equation}
and
\begin{eqnarray}
\label{zetawu}
&&W(f_1,...,f_{N})=\int
_{-\infty}^{+\infty}dz_1...dz_{N}\times \nonumber \\
&&\exp\left[-\frac{1}{g^2A}
\sum_{i=1}^{N}z_{i}^2\right]\exp\left(\frac{inz_{1}}
{2}\right)\times \nonumber\\
&&\Delta(z_1-2\pi\tilde f_1,...,z_N-2\pi f_N)\times\nonumber \\
&&\Delta(z_1+2\pi\tilde f_1,...,z_N+2\pi f_N),
\end{eqnarray}
with $$\tilde f_1=f_1+\frac{ig^2n}{8\pi}(A-2{\cal A}).$$

These formulae have a nice interpretation in terms of instantons.
Indeed, on $S^2$, there are non trivial solutions of the Yang-Mills equation,
labelled by the set of integers $f_{i}=(f_1,...,f_{N})$ 
\begin{equation}
\label{monopolo}
{\cal A}_{\mu}(x)=\left(\matrix{f_1 {\cal A}_{\mu}^{0}(x) & 0 & \ldots & 0 \cr
                         0 &\ldots&\ldots&0\cr
                         \vdots&\vdots&\ddots&\vdots\cr
                         0& 0 &\ldots &f_N{\cal A}_{\mu}^{0}(x)\cr
}\right)
\end{equation}
where ${\cal A}_{\mu}^{0}(x)={\cal A}_{\mu}^{0}(\theta, \phi)$ is the Dirac
monopole potential,
$${\cal A}_{\theta}^{0}(\theta, \phi)=0 , \quad\   {\cal A}_{\phi}^{0}
(\theta, \phi)={1-\cos \theta\over 2}, $$
$\theta$ and $\phi $ being the polar (spherical) coordinates on $S^{2}$.

The term $\exp\left[-2 \pi in f_{1}\frac{A-{\cal A}}{A}\right]$ 
in eq.(\ref{istanti}) corresponds to the classical contribution of such
field configurations to the Wilson loop.

From the above representation it is clear why the limit $A\to \infty$
should not be performed too early. Indeed $S_{inst}(f_i)$, for any finite 
set of $f_i$, goes to
zero in such a limit and fluctuations around
an instanton solution become indistinguishable from fluctuations around 
the trivial field configuration.

Only the zero instanton contribution should be obtainable 
in principle 
by means of a perturbative calculation. Therefore  
in the following we single out
the zero-instanton contribution ($f_{q}=0$, $\forall q$) in eqs.(\ref{istanti})
to the Wilson loop, obviously normalized to the zero instanton 
partition function \cite{bg}.

The equation,
after a suitable rescaling, becomes
\begin{eqnarray}
\label{zeroinst}
&&{\cal W}_{n}^{(0)}(A,{\cal A})=\frac{1}{{\cal Z}^{(0)}(A)}
\exp\left[\frac{g^2n^2
(A-2{\cal A})^2}{16 A}\right]\nonumber \\
&&\times W_{1}(0,...,0)
\end{eqnarray}
with
\begin{eqnarray}
\label{wzeroinst}
&&W_{1}(0,...,0)=\int
_{-\infty}^{+\infty}dz_1...dz_{N}\times \nonumber \\
&&\exp\left[-\frac{1}{2}
\sum_{i=1}^{N}z_{i}^2\right]\exp\left(\frac{in\sqrt {g^2A}z_{1}}
{2\sqrt 2}\right)\times \nonumber\\
&&\Delta(z_1-\frac{in}{4}\sqrt {\frac{2g^2}{A}}
(A-2{\cal A}),z_2,...,z_N)\times\nonumber \\
&&\Delta(z_1+\frac{in}{4}\sqrt{\frac{2g^2}{A}}(A-2{\cal A}),z_2,...,z_N).
\end{eqnarray}

The two Vandermonde determinants can be expressed in terms of
Hermite polynomials \cite{gross} and then expanded in the usual way.
The integrations over $z_2,...,z_{N}$ can be performed, taking
the orthogonality of the polynomials into account; we get 
\begin{eqnarray}
\label{integrata}
&&{\cal W}_{n}^{(0)}(A,{\cal A})=
\exp\left[\frac{g^2n^2
(A-2{\cal A})^2}{16 A}\right]\nonumber \\
&&\frac{1}{{\cal Z}^{(0)}(A)}
\prod_{n=0}^{N}\frac{1}{n!}\prod_{k=2}^{N}(j_{k}-1)!
\varepsilon^{j_1...j_{N}}\varepsilon_{j_1...j_{N}}\nonumber \\
&&\times\int_{-\infty}^{+\infty}dz_1
\exp\left[-\frac{1}{2}
z_{1}^2\right]
\exp\left(\frac{in\sqrt {g^2A}z_{1}}
{2\sqrt 2}\right)\nonumber\\
&&He_{j_1-1}(z_{1+})He_{j_1-1}(z_{1-}),
\end{eqnarray}
where
\begin{equation}
\label{zetapm}
z_{1\pm}=z_1\pm\frac{in}{4}\sqrt{\frac{2g^2}{A}}(A-2{\cal A}).
\end{equation}

Thanks to the relation
\begin{eqnarray}
\label{laguerre}
&&\int_{-\infty}^{+\infty}dz_1
\exp\left[-\frac{1}{2}
z_{1}^2\right]
\exp\left(\frac{in\sqrt {g^2A}z_{1}}
{2\sqrt 2}\right)\nonumber \\
&&He_{j_1-1}(z_{1+})He_{j_1-1}(z_{1-})=\nonumber \\
&&\sqrt{2\pi}(j_1-1)! \exp\bigl(-\frac{n^2g^2A}{16}\bigr)\times\nonumber \\
&&L_{j_1-1}\Bigl(\frac{n^2g^2{\cal A}(A-{\cal A})}{2A}\Bigr),
\end{eqnarray}
we obtain 
\begin{eqnarray}
\label{risultato}
&&{\cal W}_{n}^{(0)}(A,{\cal A})=
\frac 1N \,
L^{(1)}_{N-1}\Bigl(\frac{g^2 {\cal A}(A-{\cal A})\,n^2}{2A} \Bigr)\nonumber \\
&& \exp \Bigl[ -\frac{g^2 {\cal A}(A-{\cal A})\, n^2}{4A}\Bigr]  \,.
\end{eqnarray}

At this point we remark that, in the decompactification limit $A\to
\infty$, ${\cal A}$ fixed, the quantity in the equation above {\it exactly}
coincides, for any value of $N$, with eq.(\ref{risultatino}), which
was derived following completely different considerations.
We recall indeed that eq.(\ref{risultatino}) was obtained 
by a full resummation of the perturbative expansion of the Wilson loop
in terms of causal Yang-Mills propagators in LCG.
Its meaning is elucidated by
noting that it just represents the zero-instanton contribution to
the Wilson loop, a genuinely perturbative quantity \cite{bg}.

In turn it also coincides with the expression of the exact result
in the large-$n$ limit, keeping fixed the value of $n^2 {\cal A}$
(eq.(\ref{granden})). This feature can be understood if we remember
that instantons have a finite size; therefore small loops are 
essentially blind to them \cite{bgv}.
 
\section{THE GROUP ALGEBRA}

Another remarkable result follows using the relation
\begin{eqnarray}
\label{remarkable}
&&\int_{-\infty}^{+\infty}dz_1
\exp\left[-\frac{1}{2}
z_{1}^2\right]
\exp\left(\frac{in\sqrt {g^2A}z_{1}}
{2\sqrt 2}\right)\nonumber \\
&&He_{q}(z_{1+})He_{r}(z_{1-})=\nonumber \\
&&\exp\bigl(-\frac{n^2g^2}{16A}\bigl[A-2{\cal A}\bigr]^2\bigr)
\times\nonumber \\
&&\left(A-{\cal A}\right)^{\frac{q-r}{2}}{\cal A}^{\frac{r-q}{2}}\times
\nonumber \\
&&\int_{-\infty}^{+\infty}dz_1
\exp\Bigl(ingz_1\sqrt{\frac{{\cal A}(A-{\cal A})}{2A}}
\Bigr) \nonumber \\
&&\times \exp\left[-\frac{1}{2}
z_{1}^2\right]
He_{q}(z_1)He_{r}(z_1), \,\, n\ge 0.
\end{eqnarray}

Thanks to it, for $q=r=j_1-1$, eqs.(\ref{zeroinst},\ref{wzeroinst}) become
\begin{equation}
\label{nzeroinst}
{\cal W}_{n}^{(0)}(A,{\cal A})=\frac{1}{N{\cal Z}^{(0)}}\sum_{k=1}^{N}
W_{k}(0,...,0)
\end{equation}
with
\begin{eqnarray}
\label{nwzeroinst}
&&W_{k}(0,...,0)=\int
_{-\infty}^{+\infty}dz_1...dz_{N}\times \nonumber \\
&&\exp\left[-\frac{1}{2}
\sum_{i=1}^{N}z_{i}^2\right]\exp\Bigl(ingz_{k}
\sqrt{\frac{{\cal A}(A-{\cal A})}{2A}}
\Bigr) \nonumber \\
&&\Bigl[\Delta(z_1,z_2,...,z_N)\Bigr]^2.
\end{eqnarray}

This expression is nothing but the matrix
integral
\begin{eqnarray}
\label{ncrauti}
&&{\cal W}_{n}^{(0)}({A,\cal A})=\frac{1}{{\cal Z}^{(0)}}\int {\cal D}F\,
\exp(-\frac{1}{2}{\rm Tr}F^2)\nonumber \\
&&\frac{1}{N}{\rm Tr}\Bigl[\exp(igF{\cal E})\Bigr]^{n},
\end{eqnarray}
where ${\cal E}=
\sqrt{\frac{{\cal A}(A-{\cal A})}{2A}}$ 
and ${\cal D}F$ denotes the flat integration measure 
on the tangent space of constant Hermitian $N\times N$ matrices. 
 
Eqs.(\ref{nzeroinst},\ref{ncrauti}) are the deep root of eq.(\ref{crauti}). 
They
explicitly show that considering the zero instanton sector
amounts to integrating over the group algebra. Eq.(\ref{crauti})
is then recovered after decompactification ($A\to \infty$).

\smallskip

The correspondence between zero-instanton sector and
group algebra is fairly general. Another example is provided by a 
Wilson loop calculation for the adjoint representation of $SU(N)$
(for simplicity we average over different $\theta$-sectors)
\begin{eqnarray}
\label{wilsadj}
&&{\cal W}_{adj}(A,{\cal A})={1\over {\cal Z}(A)}\sum_{R,S} d_{R}d_{S}
\times \\
&&\exp\left[-{{g^2 {\cal A}}\over 4}C_2(R)-{{g^2 (A-{\cal A})}\over 4}
C_2(S)\right]\nonumber \\
&\times&\int dU \frac{1}{N^2-1}\Bigl(|{\rm Tr}U|^2-1\Bigr)
\chi_{R}(U) \chi_{S}^{\dagger}(U),\nonumber
\end{eqnarray}
$C_2(R)$ ($C_2(S)$) being now the quadratic Casimir expression of the
$SU(N)$ $R$($S$)-irreducible representation \cite{canad}. 

By repeating the expansion in terms of Young tableaux we get the 
analog of eq.(\ref{wilsonp})
\begin{eqnarray}
\label{wiltab}
&&{\cal W}_{adj}(A,{\cal A})=\frac{1}{N+1}\Bigl(1+\frac{N}{{\cal Z}(A)}
\int_0^{2\pi}d\alpha 
\nonumber \\
&&\sum_{m_i=-\infty}^{+\infty} \exp\Bigl[-\alpha+\frac{2\pi m}{N}\Bigr]^2
\Delta(m_1,...,m_N) \times\nonumber \\
&&\Delta(m_1+1,m_2-1,.,m_N)\exp\Bigl[-\frac{g^2A}{4}C(m_{j})
\Bigr]\nonumber\\
&&\times\exp\Bigl[-\frac{g^2(A-{\cal A})}{2}(m_{1}-m_{2}+1)\Bigr]\Bigr),
\end{eqnarray}
where $m=\sum_q m_q$ and $$C(m_j)=\sum_{j=1}^N\Bigl(m_j-\frac{m}{N}\Bigr)^2.$$

After a Poisson resummation, eq.(\ref{wiltab}) becomes
\begin{eqnarray}
\label{wilpois}
&&{\cal W}_{adj}(A,{\cal A})=\frac{1}{N+1}+\frac{N\exp\Bigl[\frac{g^2
(A-2{\cal A})^2}{8A}\Bigr]}{{\cal Z}(A)\,(N+1)}
\nonumber \\
&&\times \sum_{f_i=-\infty}^{+\infty}\delta(\sum_{j=1}^N
f_{j})\exp\Bigl(-\frac{4\pi^2}{g^2A}\sum_{j=1}^N
f_{j}^2\Bigr) \times \nonumber \\
&&\exp\Bigl[2\pi i(f_2-f_1)\frac{A-{\cal A}}{A}\Bigr]
\int_{-\infty}^{+\infty}dz_1\cdots dz_N\nonumber \\
&&\times \exp\Bigl[-\frac{1}{g^2A}\sum_{j=1}^{N}z_j^2\Bigr]
\exp\Bigl[\frac{i}{2}(z_1-z_2)\Bigr]\\
&\times&\!\!\!\!\!\!\Delta(z_1-\tilde f_1,z_2-\tilde f_2,
z_3-2\pi f_3,...,z_N-2\pi f_N)\nonumber\\
&\times&\!\!\!\!\!\!\Delta(z_1+\tilde f_1,z_2+
\tilde f_2,z_3+2\pi f_3,...,z_N+2\pi f_N),\nonumber
\end{eqnarray}
where $\tilde f_1=2\pi f_1+\frac{ig^2(A-2{\cal A})}{4}$ and
$\tilde f_2=2\pi f_2-\frac{ig^2(A-2{\cal A})}{4}.$

The zero-instanton contribution is again obtained by setting $f_q=0$, 
$\forall q$. We rescale the variables,
express the Vandermonde determinants in terms of Hermite
polynomials and then expand them using the completely
antisymmetric tensor. We integrate over $z_3,\cdots,z_N$ taking 
orthogonality into account. We are left with the expression
\begin{eqnarray}
\label{deter}
&&\varepsilon^{j_1,j_2,j_3,\cdots,j_N}\varepsilon_{q_1,q_2,j_3,\cdots,j_N}
(-1)^{j_2-q_2}\nonumber \\
&&\int_{-\infty}^{+\infty}dz_1
\exp\left[-\frac{1}{2}
z_{1}^2\right]
\exp\left(\frac{i\sqrt {g^2A}z_{1}}
{2\sqrt 2}\right)\nonumber \\
&&He_{j_1}(z_{1+})He_{q_1}(z_{1-})\times\nonumber \\
&&\int_{-\infty}^{+\infty}dz_2
\exp\left[-\frac{1}{2}
z_{2}^2\right]
\exp\left(\frac{i\sqrt {g^2A}z_{2}}
{2\sqrt 2}\right)\nonumber \\
&&He_{j_2}(z_{2+})He_{q_2}(z_{2-}).
\end{eqnarray}
Thanks to eq.(\ref{remarkable}) with $n=1$,
the power factors cancel owing to the antisymmetric tensors
and eq.(\ref{wilpois}) for
$f_q=0$ becomes
\begin{eqnarray}
\label{wilzero}
&&{\cal W}_{adj}^{(0)}(A,{\cal A})=\frac{1}{N+1}+\frac{N}{(N+1)\,{\cal Z}^
{(0)}}
\nonumber \\
&&\int_{-\infty}^{+\infty}dz_1\cdot dz_N
\exp\Bigl[-\frac{1}{2}\sum_{j=1}^{N}z_j^2\Bigr]\times\nonumber \\
&&\exp\Bigl[ig(z_1-z_2)
\sqrt\frac{{\cal A}(A-{\cal A})}{2A}\Bigr]\times \nonumber \\
&&\Bigl(\Delta(z_1,z_2,z_3,...,z_N)\Bigr)^2 =\nonumber\\
&&=\frac{1}{{\cal Z}^{(0)}}\int {\cal D}F\,
\exp(-\frac{1}{2}{\rm Tr}F^2)\times\nonumber \\
&&\frac{1}{N^2-1}\Bigl(|{\rm Tr}\Bigl[\exp(igF{\cal E})\Bigr]|^2-1\Bigr),
\end{eqnarray}
to be compared with eq.(\ref{wilsadj}). The matrix $F$ is
{\it traceless} in this case.

\section{CONCLUSIONS}

We conclude that, in the equal-time formulation at $D=2,$
the  area-law exponentiation (eq.(\ref{area})) 
follows, after
decompactification, only by
resumming all instanton sectors, a procedure
which changes completely the zero
sector behaviour and, in particular, the value of the string tension.

In the equal-time scenario instantons are responsible of the 
restoration of unitarity, which was threatened by the presence of ghosts.

In the light of these considerations, there is no contradiction between
the use of the causal prescription in the light-cone propagator and the
exponentiation of eq.(\ref{area}); 
this prescription is correct but the ensuing
perturbative calculation can only provide us with the expression 
for ${\cal W}^{(0)}$. 

We find quite remarkable that {\it both} quantities in eqs.(\ref{krauth})
and (\ref{area}) are (different) {\it analytic} 
functions of $g^2$. This is hardly surprising for eq.(\ref{krauth}), but
not for eq.(\ref{area}), if it is thought as a sum over instanton 
contributions. Its analytic behaviour is at the root of the possibility
of obtaining it by resumming a perturbative series
in the light-front approach.

The result above corroborates a long-standing belief, namely that
the light-front vacuum has a much simpler structure than the one
in equal-time quantization. In two dimensions such a conjecture
receives a precise {\it quantitative} support.

On the other hand the full Poisson resummation of non-analytic quantities
(the instanton contributions) has to produce the analytic expression
one expects for the Wilson loop in the heat kernel representation.
The zero-instanton sector instead is analytic on its own and, we have
shown, can be obtained by integrating over the group algebra.

We should perhaps conclude with a comment concerning higher dimensional
cases, in particular $D=4$.

We believe that most of our results are typical of the two-dimensional case.
Perturbation theory at $D=2+\epsilon$ is discontinuous in the limit
$\epsilon \to 0$; on the other hand the invariance under 
area-preserving diffeomorphism is lost when $D>2$. In a perturbative
picture the presence of massless ``transverse'' degrees of freedom 
(the ``gluons'') forces a causal
behaviour upon the relevant Green functions, whereas in the soft (IR) limit
they get mixed with the
vacuum. The light-front vacuum, which also in two dimensions is far from
being
trivial, in higher dimensions 
is likely to be simpler only as far as topological degrees of
freedom are concerned. Of course there is no reason why it should coincide
with the physical vacuum since, after confinement, the spectrum
is likely to contain only massive excitations.  
Moreover, to be realistic, ``matter'' should be introduced, both in
the fundamental and in the adjoint representation. 
Although many papers have appeared to this regard in the recent literature,
we feel that further work is still needed to reach
a satisfactory understanding.

\end{document}